\documentclass[10pt,twocolumn,letterpaper]{article}
\usepackage[table]{xcolor}
\colorlet{lightgrey}{lightgray}

\usepackage{iccv}
\usepackage{times}
\usepackage{epsfig}
\usepackage{graphicx}
\usepackage{amsmath}
\usepackage{amssymb}
\usepackage[pagebackref,breaklinks,colorlinks]{hyperref}
\usepackage{makecell}
\usepackage{pifont}
\newcommand{\cmark}{\ding{51}}
\newcommand{\xmark}{\ding{55}}
\usepackage{arydshln}
\usepackage{wrapfig}
\usepackage{booktabs}       
\usepackage{amsfonts}       
\usepackage{nicefrac}       
\usepackage{microtype}      
\usepackage{xcolor}         
\usepackage{amsmath}
\usepackage{amssymb}
\usepackage{graphicx}
\usepackage{algorithm,algpseudocode}
\usepackage{multirow}
\usepackage{multicol}
\usepackage{xcolor,colortbl}
\usepackage{tabularx}
\def\arrvline{\hfil\kern\arraycolsep\vline\kern-\arraycolsep\hfilneg}

\iccvfinalcopy 

\def\iccvPaperID{3473} 

\ificcvfinal\fi

\begin{document}

\title{DiffV2S: Diffusion-based Video-to-Speech Synthesis\\with Vision-guided Speaker Embedding}

\author{
    Jeongsoo Choi\thanks{Both authors have contributed equally to this work.} \qquad
    Joanna Hong\footnotemark[1] \qquad
    Yong Man Ro\\
    School of Electrical Engineering, KAIST\\
    {\tt\small \{jeongsoo.choi, joanna2587, ymro\}@kaist.ac.kr}
}

\maketitle
\ificcvfinal\thispagestyle{empty}\fi

\begin{abstract}
Recent research has demonstrated impressive results in video-to-speech synthesis which involves reconstructing speech solely from visual input. However, previous works have struggled to accurately synthesize speech due to a lack of sufficient guidance for the model to infer the correct content with the appropriate sound. To resolve the issue, they have adopted an extra speaker embedding as a speaking style guidance from a reference auditory information. Nevertheless, it is not always possible to obtain the audio information from the corresponding video input, especially during the inference time. In this paper, we present a novel vision-guided speaker embedding extractor using a self-supervised pre-trained model and prompt tuning technique. In doing so, the rich speaker embedding information can be produced solely from input visual information, and the extra audio information is not necessary during the inference time. Using the extracted vision-guided speaker embedding representations, we further develop a diffusion-based video-to-speech synthesis model, so called DiffV2S, conditioned on those speaker embeddings and the visual representation extracted from the input video. The proposed DiffV2S not only maintains phoneme details contained in the input video frames, but also creates a highly intelligible mel-spectrogram in which the speaker identities of the multiple speakers are all preserved. Our experimental results show that DiffV2S achieves the state-of-the-art performance compared to the previous video-to-speech synthesis technique.
\end{abstract}

\section{Introduction}
Video-to-speech synthesis techniques \cite{choi2023intelligible, hong2021speech, kim2021multi, kim2021lip, kim2023lip, schoburgcarrillodemira22_interspeech, mira2022endgan, prajwal2020learning} have been broadly studied in lip-reading research areas. It reconstructs speech from a silent talking face video, which has an advantage of not requiring extra text information of the given video input during training. However, it is still regarded as a challenging task, especially in multi-speaker and noisy environment settings, since the video-to-speech synthesis technique needs to capture the complex relationship between various lip movements and speech. 

The relationship between lip movements and speech is not always straightforward; there is considerable variation in how different people articulate sounds, as well as how their lip movements are affected by factors such as facial expressions, accents, and noise. To resolve the complicated factors that speakers themselves contain, several recent studies \cite{choi2023intelligible, kim2023lip, schoburgcarrillodemira22_interspeech, prajwal2020learning} have utilized extra speaker embedding representations obtained from the original audio information of the video input with the same speaker. The speaker embeddings are helpful for obtaining the speaker's characteristics, where those characteristics cannot be directly derived from silent talking face video. However, directly manipulating the reference audio information during the inference is not always possible as the audio information is sometimes unobtainable because of noisy environments, absence of speech, and unseen speakers during the inference time.

To alleviate the aforementioned issue, we present a novel vision-guided speaker embedding extractor using a self-supervised pre-trained model. With the largely trained audio-visual speech representation model \cite{shi2022learning}, we adopt a prompt tuning technique \cite{lester2021power} to train the certain part of the model in order to extract the appropriate speaker embedding features from the input video sequences. We set only a small amount of downstream task-specific parameters as the learnable parameters for extracting speaker embedding into the input space while freezing the other parts of the pre-trained model. By doing so, the rich speaker embedding information can be produced from solely on input visual information, and the extra audio information is not necessary during the inference time period. 

Furthermore, we propose a conditional diffusion-based video-to-speech synthesis model, named DiffV2S, using the vision-guided speaker embedding representations. As the denoising diffusion model has been proven to be effective in generating semantically meaningful representation in both image and audio processing \cite{jeong21_interspeech, lugmayr2022repaint, tevet2023human}, we also newly adopt the diffusion model to achieve high-quality mel-spectrogram containing semantically detailed information. The proposed DiffV2S is comprised of conditional diffusion modeling and sampling with speaker embedding guidance. During training, the proposed DiffV2S reconstructs a mel-spectrogram from a standard Gaussian distribution with the condition of speaker embedding representations concatenated with the visual features extracted from the input silent talking face video. During sampling, the speaker characteristics are driven to enable the model to properly adopt the speaker's style, such as voice and accent, while maintaining the articulate phoeneme details faithfully. Therefore, our model not only maintains phoneme details contained in the input video frames, but also creates a noise-free and highly intelligible mel-spectrogram in which the speaker identity characteristics are entirely preserved.

To validate the effectiveness of the proposed method, we utilize LRS2 \cite{chung2017lrs2} and LRS3 \cite{afouras2018lrs3}, the largest sentence-level audio-visual datasets obtained in the wild. Through comprehensive experiments, we show that the generated speech from the proposed DiffV2S contains much detailed contents, thus producing noise-free audio waveform with high performances.

Our key contributions are as follows:
\begin{itemize}
\item We propose a vision-guided speaker embedding extractor, so the rich speaker information can be produced solely from the input video frames. In doing so, the audio information is not necessary during the inference time.
\item We present the novel diffusion-based video-to-speech synthesis model, DiffV2S, conditioned on the speaker embedding representations and the visual representation extracted from the input talking face video. The DiffV2S not only maintains phoneme details contained in the input video frames, but also creates a highly intelligible mel-spectrogram in which the speaker identities are all preserved.
\item  To best of our knowledge, this is first time to utilize the diffusion model in video-to-speech synthesis. The generated speech from the proposed DiffV2S contains much detailed information, thus producing noise-free audio waveform with high performances.
\end{itemize}

\section{Related Works}
\subsection{Video-to-Speech Synthesis}
Video-to-speech synthesis is one of the lip-reading techniques that have been consistently studied. Ephrat and Peleg \cite{ephrat2017vid2speech} presented the initial deep-learning based video-to-speech method using an end-to-end CNN-based model. Akbari \etal \cite{akbari2018lip2audspec} utilized autoencoders in presenting a reconstruction-based video-to-speech synthesis. Prajwal \etal \cite{prajwal2020learning} introduced Lip2Wav model using a well-known sequence-to-sequence architecture to correctly capture the context.
Hong \etal \cite{hong2021speech} adopted a multi-modal memory network in video-to-speech synthesis to associate an extra audio information during inference time period. GAN-based techniques \cite{hong2022visagesyntalk,kim2021lip,mira2022endgan,vougioukas19_interspeech} were presented to produce realistic utterances from the silent talking face videos. Recent video-to-speech techniques \cite{choi2023intelligible, kim2023lip, schoburgcarrillodemira22_interspeech} utilized extra speaker embeddings from the original audio information in order to obtain the speaker's speaking styles and characteristics.
Instead of directly using the audio information, in this work, we try to extract the speaker information from the input silent talking face video sequences. To do so, we adopt prompt tuning technique.

\subsection{Prompt Tuning}
Prompt tuning refers to the process of adjusting a pre-trained model by giving it additional prompts that are relevant to a specific task or domain. With a great development of the large pre-trained language model like GPT-3 \cite{brown2020language}, prompt tuning has been drawn attention in utilizing a frozen language model for a specific task, reducing the number of model parameters for training and memory usage. Liu \etal \cite{liu2021gpt} firstly introduced P-tuning to add trainable continuous embeddings, so called continuous prompts, to the original sequence of input word embeddings. Lester \etal \cite{lester2021power} presented the modification of initial P-tuning technique by simply prepending the prompt to the input. Zhong \etal \cite{zhong2021factual} designed an effective contiuous method for optimizing prompt, so called OptiPrompt. Liu \etal \cite{liu2021p} further generalized P-tuning so that it can be comparable to fine-tuning universally across various model scales and natural language understanding tasks. While P-tuning only focuses on language models, Jia \etal \cite{jia2022visual} proposed a visual prompt tuning technique to fine-tune for large-scale Transformer models in vision. Inspired by \cite{jia2022visual, lester2021power}, in this paper, we adopt prompt tuning technique to the large pre-trained audio-visual representation model in order to obtain a proper speaker representation from the visual input.

\subsection{Diffusion Model}
Diffusion model has been spotlighted in many research areas regarding image generation. Sohl-Dickstein \etal \cite{sohl2015deep} firstly introduced diffusion probabilistic models which is a parameterized Markov chain trained using variational inference to generate samples that match the data within a specified time. Ho \etal \cite{ho2020denoising} presented progress of diffusion models that they are capable of generating high quality image samples. Along with the large usage of diffusion models in image-video generation areas, there also have been numerous studies in synthesizing audio using diffusion models. Diffusion based neural vocoders \cite{chen2021wavegrad, chen21p_interspeech, kong2021diffwave} were proposed to model the fine details of waveform conditioned on mel-spectrogram. Several text-to-speech (TTS) techniques \cite{jeong21_interspeech, liu2022diffgan, popov2021grad} also utilized the diffusion models and achieved high-fidelity and efficient speech synthesis. While the denoising diffusion model has been proven to be effective in generating semantically meaningful representation in both image and audio processing, there has not yet been explored in the video-to-speech study. Thus, it is the first time to utilize the diffusion probabilistic model in video-to-speech technique.

\begin{figure}[t]
	\begin{minipage}[b]{1.0\linewidth}
		\centering
		\centerline{\includegraphics[width=7.8cm]{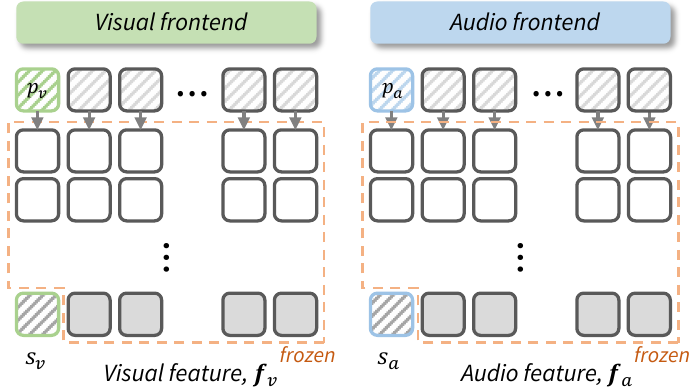}}
	\end{minipage}
	\caption{Prompt tuning via self-supervised audio-visual pre-trained model.}
	\label{fig:p}
	\vspace{-0.5cm}
\end{figure}

\section{Methodology}
Given the input talking face video sequence, $\mathbf{x}=\{x_1,\dots,x_{L}\}\in\mathbb{R}^{L\times H\times W\times C}$ where $L$, $H$, $W$, and $C$ are the frame length, height, width, and channel sizes, respectively, we design a model that synthesize the proper mel-spectrogram, $\mathbf{M}=\{M_1,\dots,M_S\}\in\mathbb{R}^{K\times S}$ with $K$ mel-spectral channel and the sequence length $S$. The main goal of this paper is to reconstruct the mel-spectrogram which contains the right styles of the input speaker as well as the articulate phoeneme details. To this end, we firstly propose a vision-guided speaker embedding extractor composed of a largely pre-trained audio-visual speech representation model modified by prompt tuning. Using the speaker embedding features, we also design a diffusion based speech synthesis model conditioned on the extracted speaker embedding. We will explain the detailed aforementioned techniques in the following subsections.

\subsection{Vision-guided Speaker Embedding Extractor}
When synthesizing the speech from a silent talking face video, it is important to know speaker's characteristics, tones, and accents to represent the proper acoustic sound. It is not always possible to extract the speaker embedding features from the audio information during the inference time due to the absence of speech and noisy environment. Thus, we design a vision-guided speaker embedding extractor that can generate the adequate speaker embedding feature without any additional audio guidance. To do so, we adopt prompt tuning \cite{lester2021power} technique which leverages few continuous learnable parameters in order to serve as prompts fed into a largely pre-trained audio-visual representation model. The prompt tuning technique is beneficial for aggregating every informative representation from the pre-trained model with adjusting the small amount of parameters for the downstream task, thus regarded as cost-effective and highly-informative.

\begin{figure}[t]
	\begin{minipage}[b]{1.0\linewidth}
		\centering
		\centerline{\includegraphics[width=8.4cm]{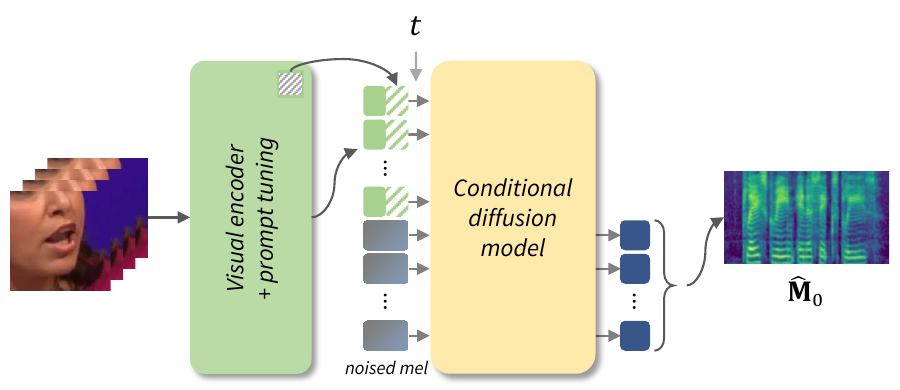}}
	\end{minipage}
	\caption{Training procedure of the proposed speaker embedding conditioned diffusion model.}
	\label{fig:1}
	\vspace{-0.3cm}
\end{figure}

\begin{figure*}[t]
	\begin{minipage}[b]{1.0\linewidth}
		\centering
		\centerline{\includegraphics[width=16cm]{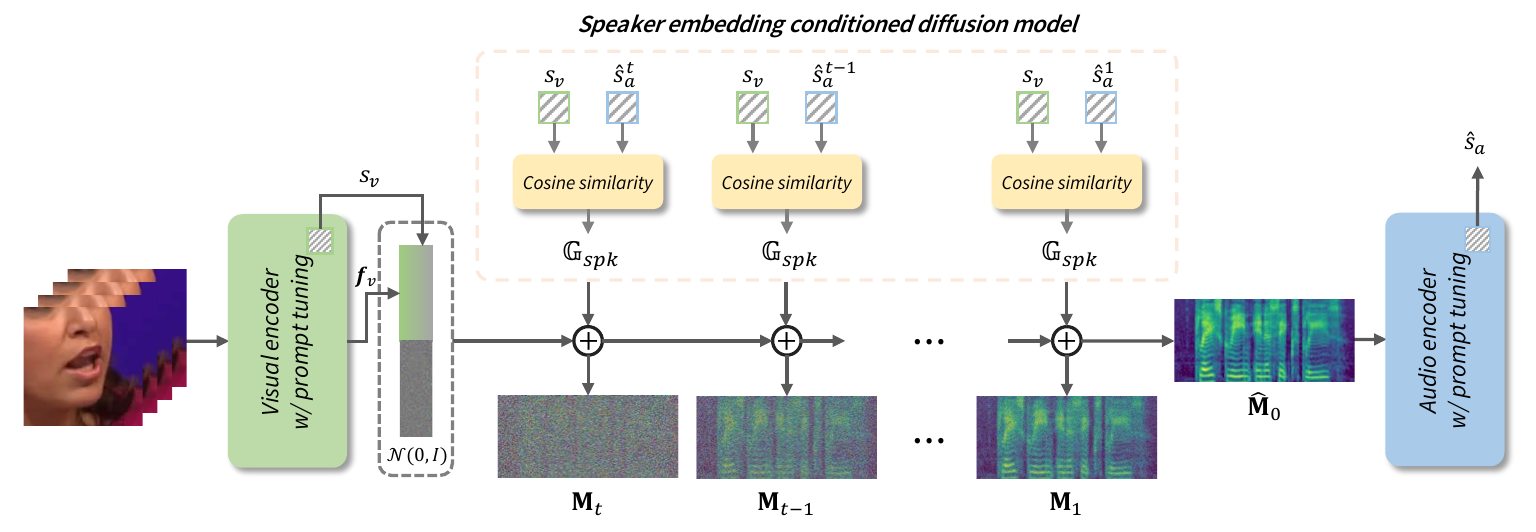}}
	\end{minipage}
 	\vspace{-0.5cm}

	\caption{Sampling procedure of the proposed speaker embedding conditioned diffusion model.}
	\label{fig:2}
	\vspace{-0.5cm}
\end{figure*}

\subsubsection{Prompt Tuning for Speaker Embedding Extractor}
Given a large pre-trained audio-visual representation model, we set $d$-dimensional learnable parameters $p_v\in\mathbb{R}^{d}$ and $p_a\in\mathbb{R}^{d}$, also called prompts, for the input video $\mathbf{x}$ and the mel $\mathbf{M}$, respectively. The prompts $p_v$ and $p_a$ are trained for extracting the speaker embeddings. To do so, we firstly extract the visual and audio embeddings, $\textit{\textbf{e}}_v\in\mathbb{R}^{L\times d}$ and $\textit{\textbf{e}}_a\in\mathbb{R}^{L\times d}$, from the visual and audio frontends, respectively:
\begin{align}
    \textit{\textbf{e}}_v &= {\mathcal{F}_v}(\mathbf{x})   \\
    \textit{\textbf{e}}_a &= {\mathcal{F}_a}(\mathbf{M}).
\end{align}
Then, each of the extracted visual and audio embeddings are taken into each pre-trained visual and audio feature extractor, $\Phi_v$ and $\Phi_a$ respectively, where the prompt is added:
\begin{align}
    [k_{v,1},\textit{\textbf{e}}_{v,1}] &= {\Phi_{v,1}}(p_v, \textit{\textbf{e}}_{v,0})   \\
    [k_{v,i+1},\textit{\textbf{e}}_{v,i+1}] &= {\Phi_{v,i+1}}(k_{v,i},\textit{\textbf{e}}_{v,i}),
\end{align}
for the $i$-th layer where $i=1, 2, \ldots, n-1$. We design the vision prompt $p_v$ to affect only the final layer $k_{v,n}$ without updating any other features through the self-attention mask. Thus, the last layer embedding $k_{v,n}$ becomes the vision-guided speaker embedding $s_v\in\mathbb{R}^{d}$, and the remainders $\textit{\textbf{e}}_{v,n}$ becomes 
$\textit{\textbf{f}}_v\in\mathbb{R}^{L\times d}$. Same procedures are applied for the audio embeddings $\textit{\textbf{e}}_a$. During the entire training phase, we only train the learnable prompts, $p_v$ and $p_a$. We keep the entire backbone pre-trained model, audio and visual encoder which of each contains the combination of the frontend and the feature extractor for the corresponding modality, remaining frozen. We also utilize the self-attention mask so that the additional prompt learns from the layer-wise feature of the pre-trained model but does not affect to the original features. The visualization of the prompt tuning technique is indicated in Figure \ref{fig:p}.

\subsubsection{Training Speaker Embedding Extractor}
\label{sec:3.1.2}
In order to train the learnable parameters, $p_v$ and $p_a$, to correctly extract the speaker embeddings, we adopt a pre-trained speaker encoder \cite{wan2018generalized} as a ground-truth guidance of speaker embeddings extracted from the actual audio input during training. It is originally trained for speaker verification task on a combination of the large speech datasets \cite{chung18b_interspeech, nagrani17_interspeech, panayotov2015librispeech}, which is trained to optimize a generalized speaker verification loss. We now call $s_{\mathcal{G}}$ as the speaker embedding guidance extracted from \cite{wan2018generalized}.

To begin with, we set each $s_v$ and $s_a$ to be similar to the speaker embedding guidance $s_{\mathcal{G}}$. Given a batch of $N$ pairs of vision-guided speaker embedding $s_v$ and the speaker embedding guidance $s_{\mathcal{G}}$, the prompt $p_v$ is trained to predict which of the $N\times N$ possible pairings across a batch actually occurred. We utilize InfoNCE loss \cite{oord2018representation} to encourage the prompts $p_v$ and $p_a$ to place the representations of positive pairs close to each other and the representations of negative pairs far apart:
\begin{align}
    \mathcal{L}_{s_v,s_{\mathcal{G}}} = -log\frac{exp(s_{v,i}\cdot s_{\mathcal{G},j})/\tau}{\sum_{k=0}^{N}{exp(s_{v,i}\cdot s_{\mathcal{G},k})/\tau}}.
\end{align}
Here, we fix the updates of the speaker embedding guidance $s_{\mathcal{G}}$. Likewise, $\mathcal{L}_{s_a,s_{\mathcal{G}}}$ can be obtained with above equation. Furthermore, we guide both $s_v$ and $s_a$ to be mapped into common embedding space through $\mathcal{L}_{s_a,s_v}$ and $\mathcal{L}_{s_v,s_a}$ so that both embeddings can relate. The final loss function of training the speaker embedding extractor is following:

\begin{align}
    \mathcal{L}_{spk\_ emb} = \mathcal{L}_{s_v,s_{\mathcal{G}}}+\mathcal{L}_{s_a,s_{\mathcal{G}}}+\mathcal{L}_{s_v,s_a}+\mathcal{L}_{s_a,s_v}.
\end{align}

Therefore, given a new video sequence or mel sequence, the speaker embedding extractor can retrieve the most plausible speaker representations from the joint embedding, even if they are not seen during training. 

\subsection{Diffusion-based Video-to-Speech Model}
\subsubsection{Training Procedure of DiffV2S}
The proposed DiffV2S generates a detailed mel-spectrogram from a standard Gaussian distribution conditioned on the channel-wise concatenation of the vision-guided speaker embeddings $s_v$ and the visual features $f_v$:
\begin{align}
c = \textit{\textbf{f}}_v || s_v.
\end{align}
Let $\mathbf{M}_1,...\mathbf{M}_T$ be a sequence of variables with $T$ number of timesteps. The forward diffusion process transforms mel-spectrogram $\mathbf{M}_0$ into a Gaussian noise $\mathbf{M}_T$ through Markov chain transitions with a predefined variance schedule $\beta_t$:
\begin{align}
  q(\mathbf{M}_t|\mathbf{M}_{t-1}, c)=\mathcal{N}(\mathbf{M}_{t};\sqrt{1-\beta_t}\mathbf{M}_{t-1}, \beta_t \mathbf{I}).
\end{align}
The reverse process is a backward of forward diffusion process, where it recovers a mel-spectrogram from a standard Gaussian noise. It can be defined as the conditional distribution and factorized into multiple transitions based on Markov chain property:
\begin{equation}
p_\theta(\mathbf{M}_{0:T},c)=p(\mathbf{M}_T,c)\prod_{t= 1}^{T}p_\theta(\mathbf{M}_{t-1}|\mathbf{M}_t,c),
\end{equation}
where 
\begin{align}\begin{split}
p_\theta(\mathbf{M}_{t-1}&|\mathbf{M}_t,c)= \\ 
&\mathcal{N}(\mathbf{M}_{t-1}; \mu_\theta (\mathbf{M}_t, t,c),\sigma_\theta(\mathbf{M}_t,t,c)\mathbf{I}).
\end{split}\end{align}
Here, $\mu_\theta$ and $\sigma_\theta$ are the mean and variance for the denoising model. In order to predict $q(\mathbf{M}_0|c)$, we need to optimize the negative log-likelihood of the predicted mel-spectrogram: $\mathbb{E}_q[\log p_{\theta}(\mathbf{M}_0|c)]$. Since $p_{\theta}(x_0|c)$ is intractable, the reparameterization trick \cite{ho2020denoising} is demonstrated to calculate the variational lower bound of the log-likelihood in a closed form. In this case, the model learns to find $\epsilon \sim \mathcal{N}(\textbf{0}, \mathbf{I})$. Nonetheless, instead of modelling $\epsilon_\theta(\mathbf{M}_t, t)$, the proposed DiffV2S is designed to predict the mel-spectrogram itself, as shown in Figure \ref{fig:1}. Thus, the diffusion model $\Psi(\mathbf{M}_t, t, c)$ predicts the mel-spectrogram $\hat{\mathbf{M}}_0$, and we utilize the L1 reconstruction loss between the predicted one and the ground truth one as follows:
\begin{equation}
{\mathcal{L}_{diff} = \mathbb{E}_{\mathbf{M}_0 \sim q(\mathbf{M}_0|c)}[\| \mathbf{M}_0 - \Psi_\theta(\mathbf{M}_t, t, c)\|_1]}.
\end{equation}

\begin{algorithm}[t!]
	\caption{Diffusion Sampling Procedure of DiffV2S with Vision-guided Speaker Embedding} \label{alg:1}
	\begin{algorithmic}[1]
		\State {\bf Inputs}: source talking face video sequence $\textbf{x}$, target mel-spectrogram  $\mathbf{M}_{\mathrm{target}}$, learnable prompts $p_v$ and $p_a$
		\State {\bf Outputs}: synthesized mel-spectrogram $\mathbf{\hat{M}}_{0}$
		\State $\mathbf{M}_T \sim \mathcal{N}(\mathbf{0}, \mathbf{I})$
		\State ${\textit{\textbf{e}}}_v \gets {\mathcal{F}}_v(\mathbf{x})$
		\State [${\mathrm{s}}_v,\textit{\textbf{f}}_v] \gets \Phi_{v}(p_v, {\textit{\textbf{e}}}_v)$
            \State $c \gets \textit{\textbf{f}}_v || s_v$
		\For {\textbf{all} $t$ from $T$ to $1$}
            \State $\widehat{\mathbf{M}}_0 \gets \Psi_{\theta}(\mathbf{M}_t, t, c) $
            \State $\widehat{\textit{\textbf{e}}}_a \gets {\mathcal{F}}_a(\widehat{\mathbf{M}}_0)$
            \State $ [{\hat{\mathrm{s}}}_a^t, \textit{\textbf{f}}_a] \gets \Phi_{a}(p_a,\widehat{\textit{\textbf{e}}}_a)$
            \State $\mathbb{G}_{spk} \gets (1-\mathrm{sim}(\mathrm{s}_v, \mathrm{\hat{s}}_a^t))$
            \State $\hat{\epsilon} \gets \frac{{\mathbf{M}}_{t} - \sqrt{\bar{\alpha}_t} \hat{\mathbf{M}}_0}{\sqrt{1-\bar{\alpha}_t}} -\sqrt{1-\bar{\alpha}_t}\nabla_{\mathbf{M}_t}\lambda\mathbb{G}_{spk}$
            \State ${\mathbf{M}}_{t-1} \gets \sqrt{\bar{\alpha}_{t-1}} \left( \frac{{\mathbf{M}}_{t} - \sqrt{1-\bar{\alpha}_t} \hat{\epsilon}}{\sqrt{\bar{\alpha}_t}} \right) + \sqrt{1-\bar{\alpha}_{t-1}} \hat{\epsilon}$
        \EndFor    
    \State \Return $\mathbf{M}_{0}$
	\end{algorithmic} 
\end{algorithm}

\vspace{-0.5cm}
\subsubsection{Conditional Sampling for DiffV2S}
During the sampling procedure, 
we utilize the vision-guided speaker embeddings $s_v$ to incorporate the gradients from the guidance of speaker embedding representations as a condition in order to guide the model to sample the mel-spectrogram with the desired speaker characteristics. To do so, inspired by \cite{kim2022diffface}, we firstly formulate the cosine similarity loss between the vision-guided speaker embedding $s_v$ and the audio-guided speaker embedding $\hat{s}_a$ extracted from the predicted mel-spectrogram $\hat{\mathbf{M}}_0$:
\begin{align} 
    [{\mathrm{s}}_v,\textit{\textbf{f}}_v] &= \Phi_{v}(p_v,{\mathcal{F}_v}(\mathbf{x})), \\
    [\hat{\mathrm{s}}_a,\hat{\textit{\textbf{f}}_a}] &= \Phi_{v}(p_a,{\mathcal{F}_a}(\hat{\mathbf{M}}_0)), \\
    \mathbb{G}_{spk} &= 1 - \mathrm{sim}(s_{v},\hat{s}_{a}),
\end{align}
where $\mathrm{sim}$ corresponds to cosine similarity. 

Using the gradients of $\mathbb{G}_{spk}$ with respect to $\mathbf{M}_t$, it is possible to derive the conditional sampling with the score function following \cite{dhariwal2021diffusion}. We utilize the deterministic sampling method DDIM \cite{song2021denoising} for enhancing the sampling speed. We derive the epsilon prediction first; then, we calculate $\mathbf{M}_{t-1}$ as follows:
\begin{equation}
     \hat{\epsilon}=\frac{{\mathbf{M}}_{t} - \sqrt{\bar{\alpha}_t} \hat{\mathbf{M}}_0}{\sqrt{1-\bar{\alpha}_t}} -\sqrt{1-\bar{\alpha}_t}\nabla_{\mathbf{M}_t}\lambda\mathbb{G}_{spk},
\end{equation}
\begin{equation}
\small
    {\mathbf{M}}_{t-1}=\sqrt{\bar{\alpha}_{t-1}} \left( \frac{{\mathbf{M}}_{t} - \sqrt{1-\bar{\alpha}_t} \hat{\epsilon}}{\sqrt{\bar{\alpha}_t}} \right) + \sqrt{1-\bar{\alpha}_{t-1}} \hat{\epsilon},
\end{equation}
where $\alpha_t=1-\beta_t$ and $\bar{\alpha_t}=\prod_{s=1}^t\alpha_s$. Algorithm \ref{alg:1} presents the more details of sampling algorithm.

\section{Experimental Setup}
\subsection{Datasets}
\noindent {\bf LRS2-BBC} \cite{chung2017lrs2} is an English sentence-level audio-visual dataset, collected from BBC television shows. It contains more than 2000 hours of videos, where both pre-train and train sets consist of about 142,000 utterances, validation set includes about 1,100 utterances, and test set contains 1,200 utterances. We utilize both the pre-training and training sets for the training, and the test set for the inference.
\vspace{0.1cm}

\noindent {\bf LRS3-TED} \cite{afouras2018lrs3} is also an audio-visual dataset in English, collected from TED and TEDx videos. It includes unconstrained long sentences with more than 50,000 vocabularies and thousands of speakers. It contains about 150,000 videos which are total about 439 hours long. About 131,000 utterances are utilized for training, and about 1,300 utterances are used for testing. We follow the unseen data splits of \cite{schoburgcarrillodemira22_interspeech}.

\begin{table*}[t!]
\centering
\renewcommand{\arraystretch}{1.1}
	\renewcommand{\tabcolsep}{1.7mm}
\resizebox{0.9\linewidth}{!}{
\begin{tabular}{c ccccc ccccc}
\multirow{3}{*}{} & \multicolumn{5}{c}{{\textbf{{LRS2-BBC}}}} & \multicolumn{5}{c}
{{\textbf{{LRS3-TED}}}} \\ \Xhline{3\arrayrulewidth} 
\multirow{2}{*}{\textbf{Method}} & \multicolumn{2}{c}{\small\textit{Low-level}} & \multicolumn{2}{c}{\small\textit{Synchronization}} & \multicolumn{1}{c}{\small\textit{Content}} & \multicolumn{2}{c}{\small\textit{Low-level}} & \multicolumn{2}{c}{\small\textit{Synchronization}} & \multicolumn{1}{c}{\small\textit{Content}} \\ 
                        & \textbf{ESTOI}$\:\uparrow$ & \textbf{MCD}$\:\downarrow$ & \textbf{LSE-C}$\:\uparrow$ & \textbf{LSE-D}$\:\downarrow$ & \textbf{WER}$\:\downarrow$ & \textbf{ESTOI}$\:\uparrow$ & \textbf{MCD}$\:\downarrow$ & \textbf{LSE-C}$\:\uparrow$ & \textbf{LSE-D}$\:\downarrow$ & \textbf{WER}$\:\downarrow$ \\ \cmidrule(l{2pt}r{2pt}){1-6} \cmidrule(l{2pt}r{2pt}){7-11}
\multicolumn{6}{l}{\textbf{\textit{with speaker embedding from audio}}}\\
SVTS \cite{schoburgcarrillodemira22_interspeech}
& 0.331 & \textbf{ 6.86 }& 7.80 & 6.47 &  71.4\% 
& 0.271 &  \textbf{8.02} & 6.04 & 8.28 &  78.0\% 
\\
Multi-task \cite{kim2023lip}
& \textbf{0.341 }&  9.37 & 6.88 & 7.32 &  57.8\% 
& 0.268 &  9.89 & 5.19 & 8.89 &  65.8\% 
\\ \cmidrule(l{2pt}r{2pt}){1-6} \cmidrule(l{2pt}r{2pt}){7-11}
\multicolumn{6}{l}{\textbf{\textit{without speaker embedding from audio}}}\\
VCA-GAN \cite{kim2021lip}
& 0.134 &  9.35 & 2.63 & 11.61 & 101.1\% 
& 0.207 &  8.85 & 4.54 & 9.63 &  95.9\% 
\\
SVTS \cite{schoburgcarrillodemira22_interspeech}
& 0.301 &  7.97 & \textbf{7.87} & \textbf{6.30 }&  76.6\% 
& 0.244 &  8.60 & 7.08 & \textbf{7.04} &  81.9\% 
\\
Multi-task \cite{kim2023lip}
& 0.322 & 10.22 & 7.19 & 7.01 &  61.0\% 
& 0.240 & 10.16 & 4.85 & 9.15 &  74.8\% 
\\
\cmidrule(l{2pt}r{2pt}){1-6} \cmidrule(l{2pt}r{2pt}){7-11}
\textbf{Proposed model}
& 0.283 & 9.85 & 7.51 & 6.90 & \textbf{52.7\% }     
& \textbf{0.284} & 9.35 & \textbf{7.28 }& 7.27 & \textbf{39.2\% } \\ \Xhline{3\arrayrulewidth}
\end{tabular}}
\vspace{0.2cm}
\caption{Performance comparisons on LRS2 and LRS3 datasets. $\uparrow$ means that the higher is the better, and $\downarrow$ means that the lower is the better.}
\label{table:1}
\vspace{-0.3cm}
\end{table*}

\subsection{Implementation Details}
\subsubsection{Data Preprocessing}
For every video, we crop based on the lip-centered region and resize the image into $88\times88$. 
We convert the mel-spectrogram from 16kHz audio using the filter size of 640 and hop size of 160 with 80 mel bands which becomes a sampling rate of 100Hz. In order to match the length of the audio feature with the visual feature of 25Hz, 40ms of mel-spectrogram is stacked to make an audio feature with dimension of 320. The mel-spectrogram is converted into a log scale and normalized into $[-1, 1]$ before entering our model.

\subsubsection{Architectural Details}
For the visual encoder, the visual features with dimension of 1024 are obtained by using publicly available \textsc{Large} AV-HuBERT \cite{shi2022learning}, and the speaker embedding is obtained through prompt tuning with the form of a vector with 1024 dimension followed by the linear layer becoming a size of 256. The speaker embedding is repeated in every frame to match the length of the visual feature sequence. We use a linear layer to project the concatenated visual features and speaker embedding to dimension of 512. The same procedure applies for the audio encoder and prompt tuning for audio representation.
During training the vision-guided speaker embedding extractor through prompt tuning technique, we utilize the self-attention mask so that the additional prompt learns from the layer-wise feature of the pre-trained model but does not affect to the original features.
For the diffusion model, we adopt Transformer \cite{vaswani2017attention} encoder architecture as a training network. We use 8 Transformer layers with 4 attention heads, hidden dimension of 512, and feed-forward layer dimension of 1024. The model uses GELU \cite{hendrycks2016gaussian} activation. The sinusoidal positional embeddings are added to each audio and visual features which means temporally synced audiovisual feature share the same positional embedding. All the input features are linearly projected into 512 dimensions before taken into the diffusion model.
Lastly, we utilize HiFi-GAN \cite{kong2020hifi} neural vocoder to convert sampled mel-spectrogram into the acutal waveform output.

\subsubsection{Training Details}
For training, we use AdamW \cite{kingma2014adam} optimizer with learning rate of $10^{-4}$. We set 1000 timesteps ($T=1000$) as default for sampling the diffusion backward process and $\lambda=1000$ to determine the guidance level of the input during sampling. We train DiffV2S for 300k updates on a GPU with batch size of 64. For computing, we use a single A6000 GPU. Note that for VCA-GAN \cite{kim2021lip}, SVTS \cite{schoburgcarrillodemira22_interspeech} (LRS3 dataset only), and Multi-task \cite{kim2023lip}, we are provided with the test audio samples from the authors. Otherwise, we re-implement and train the previous works, generate the test audio samples, and evaluate the performances.

\subsection{Evaluation Metrics}
To measure the low-level quality of the generated waveform, we utilize Extended STOI (ESTOI) \cite{estoi} as a measurement of the intelligibility of the generated speech and Mel-cepstral distortion (MCD) \cite{MCD} which quantifies the distance between the generated audio signal and the ground truth audio signal with mel-frequency cepstrum, focusing more on details. In Addition, we verify the synchronization compared to the ground truth speech. We adopt SyncNet \cite{chung2017out} and predict the temporal distance between audio and video (LSE-D) and the prediction’s average confidence (LSE-C). We utilize Word Error Rate (WER) in order to evaluate the content quality of the generated speech. We also measure the Speaker Encoder Cosine Similarity (SECS) \cite{casanova21b_interspeech} between speaker embeddings of the audio samples extracted from the speaker encoder \cite{wan2018generalized} in order to verify how much the generated speech and the original speech have similar speaker voice. 
Lastly, we conduct a human subjective study through mean opinion scores (MOS) of naturalness, intelligibility, and voice matching of the generated speech. Note that we focus on WER, SECS, and MOS metrics to examine whether the proposed model generates the clean speech samples with the right content and voice.

\begin{table}[t!]
\centering
	\renewcommand{\arraystretch}{1.2}
	\renewcommand{\tabcolsep}{1.3mm}
\resizebox{1.0\linewidth}{!}{
\begin{tabular}{cccc}
\Xhline{3\arrayrulewidth}
\textbf{Method}          & \textbf{Naturalness} & \textbf{Intelligibility} & \textbf{Voice matching}\\ \hline
\multicolumn{4}{l}{\textbf{\textit{with spk-emb from audio}}}\\
SVTS \cite{schoburgcarrillodemira22_interspeech}  & 2.16$_{\pm0.24}$ & 2.50$_{\pm0.28}$ & 2.15$_{\pm0.25}$\\
Multi-task \cite{kim2023lip}     & 1.83$_{\pm0.28}$   & 2.37$_{\pm0.33}$ & 1.96$_{\pm0.31}$  \\ \hline
\multicolumn{4}{l}{\textbf{\textit{without spk-emb from audio}}}\\
VCA-GAN \cite{kim2021lip}     & 1.35$_{\pm0.12}$ & 1.76$_{\pm0.31}$ & 1.39$_{\pm0.15}$  \\
SVTS \cite{schoburgcarrillodemira22_interspeech}  & 1.80$_{\pm0.25}$ & 2.35$_{\pm0.27}$ & 1.74$_{\pm0.21}$\\
Multi-task \cite{kim2023lip} & 1.53$_{\pm0.18}$  & 2.19$_{\pm0.34}$ & 1.60$_{\pm0.21}$\\ 
\textbf{Proposed model}  &\textbf{4.68}$_{\pm0.18}$ & \textbf{3.59}$_{\pm0.18}$  & \textbf{3.91}$_{\pm0.27}$ 
\\ \hline
Actual Voice    & 4.98$_{\pm0.02}$  & 4.93$_{\pm0.03}$ & - \\ \Xhline{3\arrayrulewidth}
\end{tabular}}
\caption{MOS comparison of the audio samples of LRS3 dataset.}
\vspace{-0.4cm}

\label{table:2}
\end{table}

\begin{figure*}[t!]
	\begin{minipage}[b]{1.0\linewidth}
		\centering
		\centerline{\includegraphics[width=16.3cm]{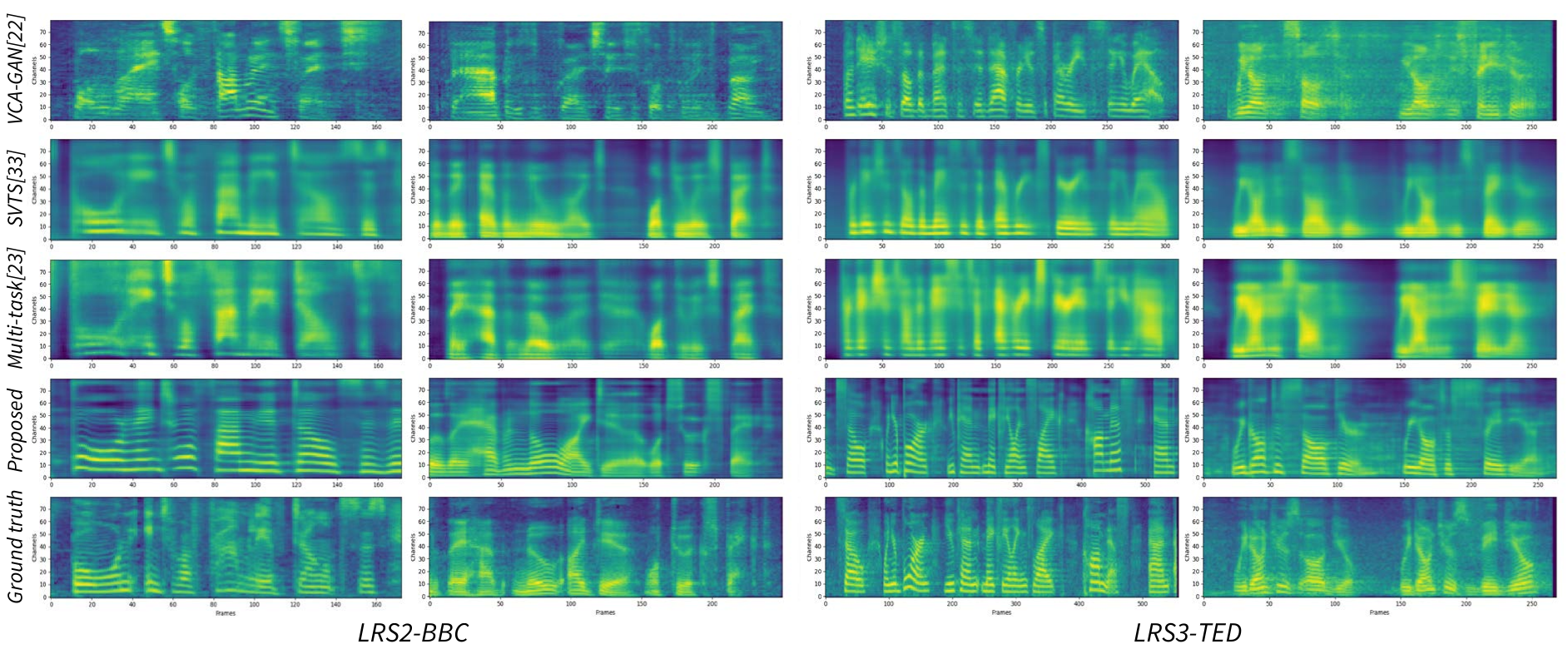}}
	\end{minipage}
 \vspace{-0.35cm}
 \caption{The sample mel-spectrogram visualizations on LRS2-BBC and LRS3-TED datasets from the previous methods \cite{kim2021lip, kim2023lip, schoburgcarrillodemira22_interspeech}, the proposed model, and the ground truth.}
\vspace{-0.35cm}
\label{fig:4}
\end{figure*}

\section{Experimental Results}
\subsection{Quantitative Results}
\subsubsection{Comparisons with the State-of-the-Arts} To begin with, we compare the performances with the previous methods \cite{kim2021lip, kim2023lip, schoburgcarrillodemira22_interspeech} using LRS2-BBC and LRS3-TED datasets. Since SVTS \cite{schoburgcarrillodemira22_interspeech} and Multi-task \cite{kim2023lip} utilize the audio guidance for extracting the speaker embedding, we reproduce both framework disregarding the audio guidance during the inference time and name \textit{without speaker embedding from audio} following the name of the work, as shown in Table \ref{table:1}. Note that we also name the future tables in the following way.

The proposed model well synchronizes with the actual input video, obtaining $7.51$ LSE-C and $6.90$ LSE-D on LRS2 and $7.28$ LSE-C and $7.27$ LSE-D on LRS3 dataset. On the low-level quality criteria, the model attains $0.283$ ESTOI and $9.85$ MCD on LRS2 and achieves $0.284$ ESTOI and $9.35$ MCD for LRS3 dataset, which are comparable scores with other methods. Most importantly, the DiffV2S considerably outperforms in regarding content quality part, achieving $52.3\%$ and $39.2\%$ on LRS2 and LRS3, respectively, which are $5.1\%$ and $26.6\%$ WER performance gaps, on both LRS2 and LRS3 respectively, compared to Multi-task \cite{kim2023lip} which best performs on content quality. Note that we focus more on the content quality extracted from the output generated waveform. Therefore, we can infer that the detailed generated mel-spectrogram also lead to the audio speech reconstruction which contain clear and right content information. 

\vspace{-0.2cm}
\subsubsection{Human Subjective Study} 
\vspace{-0.05cm}
In order to evaluate the generated speech quality, we additionally conduct a human subjective study through mean opinion scores (MOS) via three objective metrics: \textit{naturalness}, \textit{intelligibility}, and \textit{voice matching}. For the \textit{naturalness}, it verifies whether the generated speech is natural, similar to actual human speech, and the \textit{intelligibility} evaluates whether the words in the generated speech clearly sound compared to the actual transcription. Lastly, the \textit{voice matching} determines how well the results of the proposed model matches the voice of the target speaker. We use 20 randomly sampled audio from the test dataset of LRS3-TED, where each sample is evaluated by 15 participants in a 5-point scale with 0.5 increment. We also measure the 95\% confidence intervals of the total score of the participants.

\begin{table}[t!]
\centering
	\renewcommand{\arraystretch}{1.2}
	\renewcommand{\tabcolsep}{5mm}
\resizebox{0.75\linewidth}{!}{
\begin{tabular}{ccc}
\Xhline{3\arrayrulewidth}
\textbf{Method}          & \textbf{LRS2} & \textbf{LRS3} \\ \hline
\multicolumn{3}{l}{\textbf{\textit{with spk-emb from audio}}}\\
SVTS \cite{schoburgcarrillodemira22_interspeech}  & 0.558 & 0.623 \\
Multi-task \cite{kim2023lip}     & 0.525 & 0.549 \\ \hline
\multicolumn{3}{l}{\textbf{\textit{without spk-emb from audio}}}\\
VCA-GAN \cite{kim2021lip} & 0.453 & 0.445  \\
SVTS \cite{schoburgcarrillodemira22_interspeech}  & 0.499 & 0.543 \\
Multi-task \cite{kim2023lip} & 0.457 & 0.495 \\ \hline
\textbf{Proposed model}  & \textbf{0.581} & \textbf{0.625} \\ \Xhline{3\arrayrulewidth}
\end{tabular}}
\vspace{0.3cm}
\caption{Speaker Encoder Cosine Similarity (SECS) between the speaker embeddings of the generated audio samples and the ground truth audio samples.}
\vspace{-0.5cm}
\label{table:3}
\end{table}

\begin{table*}[]
	\renewcommand{\arraystretch}{1.3}
	\renewcommand{\tabcolsep}{3.0mm}
\centering
\resizebox{0.90\linewidth}{!}{
\begin{tabular}{ccccccccc}
\Xhline{3\arrayrulewidth}
 \textbf{Baseline} & \makecell{\textbf{Ground truth}\\\textbf{Spk emb}} & \makecell{\textbf{Vision-guided}\\\textbf{Spk emb}}  & \textbf{ESTOI}$\:\uparrow$ & \textbf{MCD}$\:\downarrow$ & \textbf{LSE-C}$\:\uparrow$ & \textbf{LSE-D}$\:\downarrow$  & \textbf{WER}$\:\downarrow$ & \textbf{SECS}$\:\uparrow$\\ \cmidrule(l{2pt}r{2pt}){1-3} \cmidrule(l{2pt}r{2pt}){4-9}
\cmark & \xmark & \xmark & 0.276 & 9.72 & 7.16 & 7.33 & 40.7\% & 0.608\\
\cmark & \cmark & \xmark & \textbf{0.327} & \textbf{7.85} & 7.23 & \textbf{7.27} & \textbf{38.4\% }& \textbf{0.770 }\\ \hdashline
\cmark & \xmark & \cmark & 0.284 & 9.35 & \textbf{7.28} & \textbf{7.27} & 39.2\% & 0.625\\
\Xhline{3\arrayrulewidth}
\end{tabular}}
\vspace{0.2cm}
\caption{Ablation study on LRS3 dataset analyzing the effectiveness of the vision-guided speaker embeddings. $\uparrow$ means that the higher is the better, and $\downarrow$ means that the lower is the better.}
\vspace{-0.3cm}
\label{table:4}
\end{table*}

Table \ref{table:2} indicates the MOS comparisons of the audio samples from the previous methods \cite{kim2021lip, kim2023lip, schoburgcarrillodemira22_interspeech}, the proposed method, and the ground truth audio samples from LRS3-TED dataset. The proposed model achieves $4.68$ MOS for \textit{naturalness} and $3.59$ MOS for \textit{intelligibility}. It is shown that there are large performance gaps between the MOS of the proposed framework and those of the other methods, meaning that the audio samples from the proposed model contain noise-free and detailed audio waveform. Next, for \textit{voice matching} criterion, it is clearly shown that the audio samples generated from the proposed model best follow the actual voice characteristics, achieving $3.91$ MOS. This verifies that the vision-guided speaker embedding representations are well extracted from the input videos and are adequately conditioned in the mel-spectrogram generation process, thus producing the voices similar to the ground truth voices. In addition, if the audio-guide is absent, the scores of not only \textit{voice matching} criterion but also \textit{naturalness} and \textit{intelligibility} get dropped. The means that the audio-guide actually helps the network producing proper mel-spectrogram to be similar to the actual ground truth one.

\vspace{-0.2cm}
\subsubsection{Vision-guided Speaker Embedding Analysis} 
\vspace{-0.1cm}
Lastly, we analyze the vision-guided speaker embedding to verify that the extracted speaker embedding from the vision-guided speaker extractor actually contains the correct speaker representations in comparing the ground truth ones. To do so, we calculate the cosine similarity between the extracted speaker representations from the generated mel-spectrogram and the ground truth one, respectively, which is called Speaker Encoder Cosine Similarity (SECS). Table \ref{table:3} shows that the DiffV2S achieves $0.581$ SECS and $0.625$ SECS on LRS2 and LRS3, respectively, outperforming all the previous methods, even the ones with the speaker embedding with the original audio. This clearly proves that the proposed vision-guided speaker embedding well extracts the speaker representations from the input videos. We also discover that the performances of the previous work without the actual audio input become degraded compared to those of the original ones. This shows that clearly the speaker embeddings extracted from the audio make the output speech more similar to the voice of the actual speech. Most importantly, the DiffV2S can extract the speaker embeddings by utilizing the input video with no need of extra audio information. 

\subsection{Qualitative Results}
\vspace{-0.1cm}
We visualize the mel-spectrogram samples generated from the previous works \cite{kim2021lip, kim2023lip, schoburgcarrillodemira22_interspeech} and the proposed method, along with the actual ground truth mel-spectrograms. Figure \ref{fig:4} shows two samples from each dataset, LRS2 and LRS3. It is clearly shown that the samples generated from SVTS \cite{schoburgcarrillodemira22_interspeech} and Multi-task \cite{kim2023lip} tend to be blurry and fail to bring out the details of the actual mel-spectrogram. This kind of tendency would eventually affect the waveform generation, producing noisy speech. The results from VCA-GAN \cite{kim2021lip} seem to produce the details of the mel-spectrogram; however, compared to the ground truth mels, they fail to fully follow the actual mel-spectrogram, thus producing the wrong audio speech in the end. In contrast, the generated mel-spectrogram samples from the proposed DiffV2S not only adequately represent the fine details of the mel frequency but also visually well match the ground truth mel-spectrograms. These proper and detailed mel-sepectrograms eventually lead to the speech containing the right contents without any noise. 
The demo video and the audio samples of the generated speech are available in the GitHub repository\footnote{ \url{https://github.com/joannahong/DiffV2S}}.

\subsection{Ablation Study}
We conduct the ablation study to analyze how our vision-guided speaker embeddings extracted from the input video are effective compared to the actual speaker embeddings extracted from the ground truth audio waveforms, shown in Table \ref{table:4}. The baseline refers to the visual encoder with the pre-trained model along with the vanilla diffusion model conditioned on the visual features $f_v$ only. The ground truth speaker embedding is the speaker embedding guidance $s_{\mathcal{G}}$ mentioned in Section \ref{sec:3.1.2}, which are directly extracted from the audio speech through the pre-trained speaker encoder. Lastly, the vision guided speaker embedding is the speaker representations directly extracted from the input video.

The results verify that the vision-guided speaker embeddings actually help for the overall quantitative performance, showing the performance improvement in all metrics compared to the baseline architecture. Further, the scores from the vision-guided speaker embeddings show the comparable performance with those from the actual speaker embeddings, even outperforming on several metrics, LSE-C and WER. We can infer from the results that the generated speech is affected from the vision-guided speaker embedding to contain the speaker's characteristics such as accents, so that it becomes more similar to the actual speech.

\section{Conclusion}
We propose a novel vision-guided speaker embedding extractor using the pre-trained model and prompt tuning technique. To do so, the rich speaker embedding information is produced from solely on input visual information so that extra audio information is not necessary during the inference. Using the extracted vision-guided speaker embedding representations, we propose the DiffV2S, the conditional diffusion model conditioned on the visual information and the vision-guided speaker embeddings. The DiffV2S not only saves phoneme details contained in the input video frames, but also creates a highly intelligible mel-spectrogram in which the speaker identities of the multiple speakers are all maintained. 
\vspace{0.01cm}

\noindent\textbf{Acknowledgements} This work was partially supported by the National Research Foundation of Korea (NRF) grant funded by the Korea government (MSIT) (No. NRF-2022R1A2C2005529).

{\small
\bibliographystyle{ieee_fullname}
\bibliography{main}
}

\end{document}